# Analysis of Lithiation and Delithiation Kinetics in Silicon

V.A. Sethuraman,[1,3,§] V. Srinivasan,[1] J. Newman[1,2]

[1]Environmental Energy Technologies Division
Lawrence Berkeley National Laboratory
[2]Department of Chemical Engineering, University of California
Berkeley, California 94720-8168, United States of America

Analysis of lithiation and delithiation kinetics in pulse-laser-deposited crystalline thin-film silicon (Si) electrodes is presented. Data from open-circuit relaxation experiments are used in conjunction with a model based on Tafel kinetics and double-layer capacitance to estimate the apparent transfer coefficients ($\alpha_a$, $\alpha_c$), and exchange current density to capacitance ratio ($i_0/C_{dl}$) for lithiation and delithiation reactions in a lithiated silicon ($Li_xSi$) system. Parameters estimated from data sets obtained during first-cycle amorphization of crystalline Si, as well as from cycled crystalline Si and amorphous Si thin-film electrodes do not show much variation, indicating that they are intrinsic to lithiation/delithiation in Si. A methodology to estimate the side-reaction rate and its role in the evolution of the open-circuit potential of the $Li_xSi$ system are discussed. We conclude that the large potential offset between lithiation and delithiation reactions at any given state of charge is partially caused by a large kinetic resistance (*i.e.,* small $i_0$). Using the estimated parameters, the model is shown to predict successfully the behavior of the system under galvanostatic lithiation and delithiation.

[§] Vijay_Sethuraman@Brown.edu (V.A. Sethuraman)
[3]Present address: School of Engineering, Brown University, 182 Hope Street, Providence, Rhode Island 02912, USA.

## Introduction

Electrochemical lithiation and delithiation of silicon can be ideally represented as,

$$xLi^+ + Si + xe^- \leftrightarrow Li_xSi \qquad [1]$$

The fully lithiated phase of silicon at room temperature is $Li_{3.75}Si$, which translates to a maximum theoretical capacity of 3579 mAh $g^{-1}$ for silicon,[1] much higher than that of graphite (372 mAh $g^{-1}$).[2,3] When used in a battery, this high capacity results in a significant increase in the theoretical energy density and specific energy of the cell (by as much as 25 to 30%) and could help lower the cost per kWh. This high capacity combined with silicon's low discharge potential makes it an attractive choice for use as negative electrodes in lithium ion batteries. However, the large first-cycle capacity loss, continuous side reactions during cycling, and the large volume change (*ca.* 300%) have all been detrimental to the commercialization of this system.[4] The electrochemical lithiation and delithiation of silicon at ambient temperatures has been extensively studied in the recent years in such forms as nanowires,[5-7] amorphous thin films,[5-12] crystalline thin films,[5] crystalline powder,[14,15] composites,[14,17] and mixtures with carbon.[5-20] The above-mentioned detrimental characteristics appear in all of these studies. Reference 5 reviews the methodologies adopted for reducing the capacity loss observed in silicon anodes and the challenges that remain in using silicon and silicon-based anodes.

While much progress as been made in understanding the means of enabling Si anodes to cycle reversibly, one interesting feature of this electrode is the potential offset that exists between charge and discharge. The lithiation potential for a given state of charge is considerably lower than the delithiation potential at that state of charge (SOC). Furthermore, the difference between lithiation and delithiation potentials at a given SOC, defined as the potential offset, appears to be nearly rate independent.[5,22] For example, even at low rates (*i.e.*, C/10), the potential *vs.* $Li/Li^+$ during lithiation is lower than during delithiation by approximately 0.32 V. Furthermore, data in the literature seem to indicate that particle size[23] and film thickness[24] have an effect on the potential offset, which varies from as high as 300 mV for composite electrodes[1] to less than 250 mV for nanowires[5] and amorphous thin-film electrodes.[24] As a consequence of this potential offset, the silicon electrode exhibits a stable hysteresis loop (as in a potential *vs.* capacity plot) at every SOC upon lithiation and delithiation. Similar to nickel hydroxide,[25] a hysteresis loop created during a complete lithiation and delithiation cycle is not sufficient to define the state of the $Li_xSi$ system, even at low rates. Also, the potential obtained at any SOC depends on the cycling history of the $Li_xSi$ system and therefore cannot be used as an indication of the SOC of the cell. This potential offset also results in lowering the cell efficiency (*e.g.*, 91.5% cell energy efficiency at low rates when paired with a 3.8 V $LiNi_{0.8}Co_{0.15}Al_{0.05}O_2$ cathode).

There are several documented examples of potential hysteresis in electrochemical systems such as the NiOOH electrode,[25] carbon nanotubes,[26-28] bulk graphite,[29-32] $Li_xWO_3$,[33] $Li_{1\pm y}NiO_2$,[34] $Pr_{0.7}Ca_{0.3}MnO_3$,[35-36] $LiMnO_2$,[37-39] $LiMoN_2$,[40] $Li_{3-y}V_2(PO_4)_3$,[41] $Li_ySiSnON$,[42] $LiFePO_4$,[43,44] Si-Sn alloys,[45] Si-C-O,[46] certain conducting polymers[47] such as α-phenylenes, α-thiophenes,[48] polypyrrole,[49] and polyaniline,[50,51] $La_{0.9}Sr_{0.1}MnO_3/YSZ$,[52,53] PEM fuel cells,[54] and certain redox proteins.[55] There is no single reason as to why potential hysteresis occurs in all of these systems – for example, potential hysteresis upon lithium intercalation and deintercalation in hydrogen containing carbons was attributed to lithium binding on hydrogen terminated edges of hexagonal carbon fragments resulting in a $sp^2$ to $sp^3$ bond transition.[29] A simple model which accounts for the energy associated with this bonding change was able to predict the observed potential hysteresis. This Arrhenius-type model also predicts a decrease in potential offset by 59.8 mV for every order of magnitude change in discharge rate, which suggests that the phenomenon is reversible (*i.e.,* the potential offset is rate dependent). On the other hand, in systems exhibiting phase transitions upon lithium insertion and de-insertion such as $LiMnO_2$, the potential hysteresis is thought to be caused by domain-like microstructures with spinel embedded in layered material.[37]

Potential hysteresis exhibited by amorphous Si-Sn alloys, the closest to the system studied in this work, upon lithiation and delithiation is thought to occur by differences in energy dissipated during the changes in the local atomic environment around the host atoms and lithium.[45] Neudecker *et al.*[42] measured the open-circuit potential relaxation from the lithiation and delithiation curves for various phases in a $Li_xSiSnON$ electrode in a solid-state battery (*i.e.,* without a liquid electrolyte) and found that the potentials were evolving at an extremely slow pace. They fit the potential relaxation to logarithmic time dependence and, extrapolating to 10 years, still obtained a potential offset of approximately 100 mV between the lithiation and delithiation profiles. Because of this, they attribute this behavior to a true thermodynamic hysteresis, and argued that the origin for this could be presence of metastable domains in the electrode which are sensitive to the direction of lithium transfer.[42] Similar behavior was also observed in amorphous $Li_xMn_{2-y}O_4$ cathodes, also thought to be caused by metastable domains.[39] Although potential hysteresis in electrochemical systems resembles those exhibited by ferroelectric hysteresis, the difference lies in compositional inhomogeneity typically seen in the former. One of the objectives of this study is to understand, quantify, and describe mathematically this potential offset seen in the silicon electrode.

We have recently measured *in situ* the stresses generated in the silicon thin-film electrode during lithiation/delithiation, and shown that *ca.* 40% of the energy loss can be accounted for by mechanical dissipation.[56,57] However, the origin of the rest of the potential offset remains indeterminate. In this paper, we postulate that the $Li_xSi$ system is kinetically limited at practical lithiation and delithiation rates (*i.e.,* C/30 to 3C), and that a small exchange current density ($i_0$) causes the observed phenomena. As a result, the open-circuit-potential relaxation is slow and takes a long time to reach equilibrium. The objective of this study is to explain quantitatively the electrochemical behavior of the silicon electrodes during lithiation and delithiation and to check the above-mentioned postulate. Inherent in these objectives lie answers to the following questions: (i) What causes the potential offset between lithiation and delithiation in this system, and how does this potential offset behave under various electrochemical conditions? (ii) What parameters characterize this offset, and how can they be extracted quantitatively? (iii) Can the extracted parameters then be used in conjunction with a continuum model to predict the electrochemical behavior of the system? and (iv) Can this potential offset be eliminated? These objectives were accomplished by a series of cycling and open-circuit relaxation experiments on a pulse-laser-deposited (PLD) thin-film silicon electrode in conjunction with a kinetic model based on Tafel kinetics with double-layer charging with and without the side reaction (*i.e.,* electrolyte reduction). The PLD silicon electrode was chosen as a model system because it behaves like a composite electrode made with crystalline Si nanoparticles without presenting the mathematical complexities from a distributed reaction. In addition, thin-film Si electrodes exhibit minimal capacity fade and are reversible for a moderate number of cycles. In the first part of this study, the quasi-equilibrium potential was measured as a function of SOC from a series of open-circuit relaxation experiments on a fresh silicon electrode (*i.e.,* initial lithiation/delithiation resulting in amorphization) as well as on a well-cycled electrode. The data collected from these relaxation experiments were used to estimate the apparent transfer coefficients and the ratio of exchange current density to double-layer capacitance for lithiation and delithiation reactions as a function of SOC. These kinetic parameters were then used to predict the lithiation and delithiation potential profiles. The model results prove our initial hypothesis and give insights into the origin of potential offset and ways to eliminate this potential offset.

## Experimental

*Substrate preparation.* – 1.2-mm-thick stainless steel disks (diameter = 1.8 cm) were successively wet polished with 140, 75, 23, and 12.5 μm silicon carbide abrasive papers (Leco Corp., St. Joseph, MI) and cleaned successively in ultrasonic baths of acetone, methanol, and de-ionized water for 15 minutes each. The substrates were then attached to a substrate holder in a home-built pulse-laser deposition chamber such that the polished surface was facing the plume. The distance between the substrate and the target was 5 cm.

*Electrode preparation.* – For Cu thin films, the deposition was carried out at room temperature (*ca.* 23°C) in vacuum with a base pressure of $2.66 \times 10^{-5}$ Pa. Silicon thin films were deposited in an argon atmosphere with a pressure of 66.7 Pa. Before deposition, the chamber was pumped down to a base pressure of $2.66 \times 10^{-5}$ Pa. A pulsed krypton-fluoride (KrF) excimer laser (λ = 248 nm, Lambda Physik LPX 210i) was focused onto the targets. Pure copper and silicon (99.999%, Super Conductor Materials, Inc.) targets were used to deposit copper and silicon respectively. The energy flux on the target was approximately 2.4 J/cm$^2$, and the incident angle between the laser beam and the target normal was 45 degrees. The pulse frequency was 10 Hz, and the pulse duration was 25 ns. A layer of copper was deposited followed by a layer of silicon. The copper underlayer provides better film adhesion, as shown by Maranchi *et al.*,[58] and by Sethuraman *et al.*,[59] and without this layer the cell did not cycle very well. The respective deposition times were 14 and 180 minutes.

*Electrode characterization.* – The pulse-laser-deposited (PLD) sample was examined in a high-resolution JEOL JSM-6340F field-emission scanning electron microscope operated at an accelerating voltage of 5 kV using 5 mm as the working distance with the secondary and backscattered electron-image detectors. Energy dispersive X-ray (EDX) spectroscopy was carried out using a Genesis XM2 microanalysis system (EDAX Inc., Mahwah, NJ) to evaluate the surface composition of the film. The surface was also analyzed by Raman microscopy (Labram, ISA Groupe

Horiba) with a helium-neon (HeNe) laser ($\lambda$ = 632.8 nm) at 1 mW power as the excitation source. The electrode was then assembled into a coin cell configuration (#2325, *i.e.,* 23 mm diameter and 2.5 mm total thickness) obtained from the National Research Council (NRC, Canada) with a lithium metal counter and reference electrode (diameter = 1.5 cm) and a woven Celgard 2500 separator (diameter = 1.9 cm, Celgard Inc., Charlotte, NC). 1.2 M lithium hexafluoro phosphate in 1:2 (vol. %) ethylene carbonate/diethyl carbonate (Ferro Corporation, Independence, OH) was used as the electrolyte.

Electrochemical measurements were conducted in an environmental chamber at 25°C (±1°C) using a Solartron 1480A MultiStat system (Solartron Analytical, Oak Ridge, TN), and data acquisition was done using Corrware (version 2.8d, Scribner Associates Inc., Southern Pines, NC). The cell was cycled galvanostatically at 20 $\mu$A/cm$^2$ between 1.2 and 0.01 V *vs.* Li/Li$^+$. The data acquisition rate was 1 Hz for all the electrochemical experiments. Rate experiments (from C/30 to C/4) were conducted between an upper cut-off potential of 1.2 V *vs.* Li/Li$^+$ and 50% SOC. This limit was chosen to avoid lithium plating and also to avoid the formation of the crystalline Li$_{22}$Si$_5$ phase. Open-circuit relaxation experiments were conducted as a function of SOC at 16% intervals. The input impedance of the instrument was 12 G$\Omega$, and hence the current due to the open-circuit-potential measurement was negligible.

## Results and Discussion

Figure 1 shows the surface morphology of the pulse-laser deposited (PLD) silicon film. The film exhibits a highly cracked surface with a grain-size distribution centered approximately on 140 nm and appears to be porous. The presence of bigger chunks of silicon on the surface is due to the randomness of the laser ablation of the silicon target and is typical of PLD films. The mean-crystallite size distribution in these PLD films is similar to that of Si particles used to fabricate composite electrodes.[14] A detailed study on the properties of PLD silicon thin films deposited using a KrF excimer laser can be found in the study reported by Chen *et al.*[60] The elemental X-ray analysis of the PLD film (not shown) indicates that the surface is predominantly silicon. The Raman spectrum (shown in Figure 2) obtained on the PLD silicon thin film is identical to that from a boron-doped Si (100) wafer with a sharp peak at 521 cm$^{-1}$ and is therefore highly crystalline. The mass of the film was not measured, and therefore the capacity of the film is either normalized to the maximum capacity obtained or expressed as charge (in A·s units) throughout this article.

*Film behavior.* – Figure 3 shows the potential curves of the PLD silicon thin-film electrode cycled at C/8 rate ($I_{app}$ = 20.68 $\mu$A/cm$^2$) between 1.2 and 0.01 V *vs.* Li/Li$^+$. These curves were obtained on a cell that had reached steady-state cycling efficiency (*ca.* 5 cycles). The cycling behavior of the PLD Si thin film is very similar to that of a composite electrode made with crystalline silicon powder.[14] The potential during the initial lithiation (shown in Figure 3 inset) is relatively flat due to solid-state amorphization of the crystalline Si thin film, and the corresponding first-cycle irreversible capacity loss is approximately 60%. The potential offset between charge and discharge was 320 mV at 50% SOC during steady-state cycling. The steady-state lithiation capacity was higher than that of the delithiation capacity by approximately 1.8%. Although there was no evidence for the formation of the crystalline $Li_{3.75}Si$ phase during lithiation below 50 mV *vs.* $Li/Li^+$, the delithiation potential profile also exhibits a characteristic flat plateau in the neighborhood of 50% SOC. Based on literature data and from the sloping potential profile, we believe we are cycling in the amorphous region. Figure 4 shows the lithiation and delithiation capacities for the first few cycles. Similar to other forms of silicon electrodes reported in the literature, the PLD silicon exhibits a large irreversible capacity loss during the first cycle (*ca.* 60%), and the cycling efficiency goes up above 98% for the subsequent cycles, evidence of side reactions contributing to the loss in efficiency. The large surface-to-volume ratio typical of thin films provides a large area for electrolyte reduction and SEI formation for a given electrode capacity.

Figure 5 shows data obtained between 0 and 50% SOC for different lithiation and delithiation rates (from C/30 to C/2) on this cell. It can be seen that the cycling rate has very little impact on the potential offset between lithiation and delithiation, similar to the rate capability data reported on Si nanowires[5] and amorphous Si thin films.[22] In addition, lithiation appears to be somewhat more rate dependent than delithiation. In general, for a given SOC, a potential that is invariant with current is taken to be indicative of a system that is operating at thermodynamic conditions. If this were true, the data in Figure 5 would suggest that the Si electrode exhibits two equilibrium potentials, one on lithiation and one during delithiation. We test the nature of these potentials by performing potential-relaxation experiments (Figure 6) which shows the relaxation of cell potential at open circuit at 50% SOC after lithiation and delithiation at a C/8 rate.

The open-circuit potential corresponding to delithiation (the curve on the top) decreases rapidly at first (< 2 hours), levels out (from 2 to 10 hours), and then increases slowly (> 10 hours). On the other hand, the open-circuit potential corresponding to lithiation (the curve in the bottom) increases more rapidly at first (< 2 hours) and then evolves more slowly (> 2 hours). Firstly, this indicates that the closed-circuit potential (*i.e.,* under galvanostatic conditions) is not the equilibrium potential for this system. Secondly, the fact that both curves evolve slowly towards a higher value at longer times indicates the presence of a side reaction, which makes it difficult for the potentials to collapse onto an equilibrium value for this SOC. The presence of the side reaction is not surprising considering the large overpotential for the solvent reduction reaction at these potentials. Further progress in estimating the true state of this system requires a methodology to account for the side reaction. We perform this correction in a manner similar to that shown in previous studies, as described below.[25,61,62]

*Side-reaction correction* – If the marching behavior seen from cycle to cycle (*e.g.,* in Figure 3 for a cell that has reached steady cycling) is caused by a side reaction, the applied current during the lithiation process can then be written as:

Total current $(i_{app})$ = Lithiation current $(i_{main})$ + Electrolyte reduction current $(i_{side})$     [2]

Similar to the approaches taken by Darling and Newman[61] for the $Li_yMn_2O_4$ system, and by Ta and Newman[62] for the nickel hydroxide system, we assume Tafel kinetics for the side reaction (electrolyte-reduction reaction) in our case. The current due to this reaction can be written as,

$$i_{side} = i_{0,side} \exp\left[-\frac{\alpha_{side}F}{RT}(V - U_{side})\right] \qquad [3]$$

The transfer coefficient for the side reaction, $\alpha_{side}$, was assumed to be 0.5. While Tafel kinetics does not provide an explicit equilibrium potential ($i_o$ and $U$ are related), we assume a value of $U_{side}$=0.8 vs. Li/Li$^+$ to estimate $i_0$. This side-reaction current was then calculated through the cycle assuming an $i_0$ such that the marching was eliminated from the cycling data. The steady cycling data shown in Figure 7a were corrected for the side reaction, and the result is shown in Figure 7b. The corresponding value for $i_0$ was 7.5 x 10$^{-13}$ A/cm$^2$, based on the cross-sectional area. A similar side-reaction-rate estimation was also done for the first few cycles (*i.e.*, cycles 2 to 4), and the estimated value for $i_{0,side}$ was slightly higher at 1.35 x 10$^{-12}$ A/cm$^2$. A similar decrease in capacity fade or increase in cycling efficiency numbers upon cycling is consistently seen in data reported in the literature. Using the extracted kinetic parameters, the true SOC of the system can be determined under various conditions by correcting for the side reactions.

*Equilibrium potential of the Si electrode:* Figure 8 shows the open-circuit potential after 10-hour relaxation period for lithiation and delithiation as a function of SOC. The figure shows that, similar to what is described above, the potential relaxes to a lower value from the lithiation curves compared to the delithiation curve, at the same SOC. Close examination of the data shows that the potential relaxes to 60% of the offset between lithiation and delithiation. Complete potential relaxation is not achieved even on longer open circuit times, and is essentially complicated by the presence of the side reaction, leading to the self-discharge (*i.e.*, self delithiation) of the electrode. As mentioned earlier, in our studies on stress effects in similar films,[56] 40% of the energy lost between charge and discharge is accounted for due to mechanical dissipation. We note that the potential drop between the lines and the symbols in Figure 8 at 50% SOC account for 60% of the energy loss. The correlation between these numbers suggests that stress has a role in the potentials measured in this figure. Understanding the impact of stress and mechanical dissipation on the chemical potential of silicon would require a much more detailed study of this system, which is outside the scope of this paper.

Figure 6 and 8 suggest that achieving a true equilibrium value at a given SOC would require relaxation to very long times; longer than the relaxation times reported here. However, the presence of the side reaction results in the self-discharge of the electrode continuously, further complicating this estimation. A methodology similar to the one described above has been used in the past to estimate an equilibrium potential when a side reaction is present.[25,61,62] References 25 and 62 are very similar to the present system in that the chemistry studied by these authors (the NiOOH electrode) also exhibits a potential hysteresis. However, in both these papers the authors were studying a system that was thought truly to have multiple potentials at the same SOC, and therefore no attempt was made to estimate a single equilibrium potential from the data. The Si system differs from the NiOOH system in that the closed-circuit potential does not appear to represent the equilibrium potential, suggesting that the equilibrium potential lies somewhere between the close-circuit potentials. Estimating an equilibrium potential for Li insertion in Si has been a subject of previous study. Chevrier and Dahn used first-principles calculations to estimate this potential recently and show a single curve between the closed-circuit potentials.[63] They argue that the breaking of Si bonds causes the hysteresis in voltage. On the other hand, Baggetto *et al.*, report two equilibrium curves *versus* SOC, one each for lithiation and delithiation, using a galvanostatic-intermittent-titration technique (GITT).[64] The authors argue that the two equilibrium potentials suggest a thermodynamic hysteresis; however no proof is provided for this conclusion. In other words, the authors argue that the symbols shown in Figure 8 are the true equilibrium values. As described above, the relaxation of potential in Si is a very slow process and estimating the potential from a GITT when there is a significant side reaction present is not straightforward.

*Parameter estimation from open-circuit relaxation data.* – The kinetic parameters for the $Li_xSi$ system were estimated using data obtained from the open-circuit relaxation experiments. In the mathematical treatment of the system, porous-electrode effects were ignored due to the use of a very thin film (thickness ≈ 500 nm) and the current small enough to ignore diffusion losses. For example, for diffusion coefficient values of $10^{-8}$ $cm^2/s$ reported by Yoshimura *et al.*[65] and $5.1 \times 10^{-12}$ $cm^2/s$ reported by Ding *et al.*,[66] the respective time constants are 2.5 ms and 5 s, which are small compared to the hours associated with potential relaxation seen in this system.

Similar to the approach taken by Davis *et al.*[67,68] to model the behavior of the Li-$CF_x$ system, the open-circuit potential relaxation due to the main and side reactions driven by double-layer capacitance can be written as,

$$-C_{dl}\frac{dV}{dt} = i_0\left\{\exp\left[\frac{\alpha_a F}{RT}(V-U)\right] - \exp\left[-\frac{\alpha_c F}{RT}(V-U)\right]\right\} - i_{0,side}\exp\left[-\frac{\alpha_{side}F}{RT}(V-U_{side})\right] \quad [4]$$

The dependence of stress on the kinetics of electrochemical lithiation and delithiation is ignored in writing this expression. Equation 5 is general and is used for both lithiation and delithiation. We now describe the nature of the curve qualitatively. For the lithiation reaction, assuming Tafel kinetics, for a case without a side reaction, the analytic solution for equation 5 is

$$V = U + \frac{RT}{\alpha_c F}\ln\left\{\frac{\alpha_c F i_0}{RTC_{dl}}t + \exp\left[\frac{\alpha_c F}{RT}(V_0-U)\right]\right\} \quad [5]$$

where $V_0$ is the initial potential and U is the equilibrium potential, estimated as described above. At long times, this equation reduces to,

$$\frac{dV}{d\ln t} = \frac{RT}{\alpha_c F} \quad [6]$$

corresponding to a straight line in a plot of V *vs.* ln(t). Such behavior has been shown to be exhibited by the lead acid system,[69] Li-$CF_x$ system,[67,68] and electrochemical capacitors and is considered a characteristic of a system that is limited by Tafel kinetics and a double layer capacity.[70]

The equations above can be rearranged as shown by Davis *et al.* in dimensionless form to yield two dimensionless parameters, one that involves $\alpha_c$ and the second that combines $i_o$ and U. The rearrangement highlights the fact that in Tafel kinetics, the $i_o$ and U are connected and cannot be independently determined. However, in order to simulate the voltage of the battery under closed-circuit conditions, both the $i_o$ and the U are needed. Therefore, for the purpose of aiding the model-development effort, we report both these quantities in this paper.

In this paper we take the equilibrium to be a single potential curve between the closed circuit potentials. This is shown as a dashed line in Figure 8 and is obtained by extrapolating the potential evolutions on the open circuit. Note that this choice to this potential function is arbitrary and is only needed for the simulations reported below. This potential as a function of SOC was fit to a polynomial, shown in Figure 8 and given by

$$U = -4.76z^6 + 9.34z^5 - 1.8z^4 - 7.13z^3 + 5.8z^2 - 1.94z + 0.62, \, 0 \leq z \leq 1 \quad [7]$$

The potential in the vicinity of $z = 0$ in Figure 8 is an artifact and is not the result of the open-circuit relaxation experiments. Using the U reported here, the values for the apparent transfer coefficients ($\alpha_a$, $\alpha_c$) and the ratio of the exchange current density to the double layer capacitance ($i_0/C_{dl}$) were estimated by fitting equation 5 with the experimental data. The data fits were obtained using a Levenberg-Marquardt[71] algorithm to minimize the sum of square of the error between the model prediction and the experimental data. Figure 9 shows the result for one such parameter-estimation procedure corresponding to open-circuit relaxation at 83% SOC during lithiation for $i_0/C_{dl}$ = 8.64 nV/s and $\alpha_c$ = 2.06. Model fits with and without the side reaction are shown. The former was the result of the analytic solution in equation 5, and the latter was the result of numerical solution to equation 4. The simulation show that at times less than $5 \times 10^4$ s the two simulations yield identical results, while deviations start to occur beyond this point. While the model with the side reaction shows a change in the Tafel slope, the model without the side reaction does not show this change.

This rise in potential slope at large times can be further understood by looking at the simulated current density to the main and side reactions on open circuit, as shown in Figure 10. When the electrode is operating at open circuit, the double-layer discharges onto the two reactions. The model predictions suggest that at short times (less than $5 \times 10^4$ s) the main reaction dominates while the side reaction component is lower, a consequence of the kinetics of the two reactions. However, as the potential of the electrode approaches the equilibrium potential of the main reaction, the current to the main reaction starts to decease until its magnitude is lower than that for the side reaction. Beyond this point, the side reaction kinetics dominates the nature of the electrode's behavior. Open-circuit experiments that are conducted for longer times will lead to the self discharge of the electrode until the potential reaches the equilibrium potential of the side reaction (~0.8 V *vs.* Li/Li$^+$), and the overpotential for the side reaction becomes negligible. The change in the Tafel slopes observed in Figure 9 provides a means of estimating the apparent transfer coefficient for the side reaction. However, this avenue was not pursued in this paper and the value was assumed to be 0.5.

The method outlined above was used to estimate the apparent transfer coefficients ($\alpha_a$, $\alpha_c$) and the ratio of the exchange current density to the double layer capacitance ($i_0/C_{dl}$) for various SOC for the main reaction. The exchange current density can be further split to eliminate the activity coefficient dependence by using the expression,

$$i_0 = i_0^{ref} (z)(1-z) \frac{dU}{dZ} \qquad [8]$$

the pre-exponential $i_0^{ref}$ refers to the rate constant for the lithiation or delithiation reaction without the dependence of the activity coefficient. The parameters $i_0^{ref}/C_{dl}$ and $\alpha_a, \alpha_c$ were estimated as a function of SOC, and the results are shown in Figure 9 and Figure 10, respectively. The resulting values for the parameter $i_0^{ref}/C_{dl}$ of the order of nV/s indicates a very large time constant for equilibration in this system (~1 month). The estimated values for $i_0/C_{dl}$ was relatively constant with the SOC. Note that the estimated values of $i_0/C_{dl}$ at any given state of charge depend on the value of the equilibrium potential at that state of charge. For example, if the potentials at the end of the 10-hour relaxation period at 50% state-of-charge were used as equilibrium potentials, the estimated values of $i_0/C_{dl}$ are larger by an order of magnitude than those reported in Figure 9. The apparent transfer coefficient was fit to a linear relation *versus* the SOC as given below

$$\begin{aligned}\alpha_a &= 1.77z + 1.65 \\ \alpha_c &= 0.63z + 1.52\end{aligned} \qquad [9]$$

These values agree very well with those estimated by Baggetto *et al.* (see figure 9 in reference 64). The values of $\alpha_a$ and $\alpha_c$ are significantly higher than 0.5 indicating complex lithiation and delithiation reactions. Very large apparent transfer coefficients have been observed in some cases when the reactant is strongly solvated in a polar solvent.[72] In addition, as was pointed out before, stress effects[56] could also play a role during the open-circuit-potential behavior, modifying the nature of the potential evolution and contributing to the estimation of the apparent transfer coefficient as calculated in this study. A detailed model that incorporates the effect of stress would further help in clarifying this issue.

The open-circuit-potential relaxation experiments were repeated during the first-cycle lithiation and delithiation in a PLD Si thin-film electrode as well as in amorphous Si thin-film electrodes. A 10-hour open-circuit-potential relaxation experiment at various SOC intervals during the first lithiation of a PLD Si thin-film electrode is shown in Figure 11. The flat potential profile is a characteristic of solid-state amorphization of crystalline silicon upon lithiation.[12] The inset in this figure shows the open-circuit potential evolution at 50% SOC along with the model fit based on equation 7. Between lithiation on a fresh PLD silicon thin-film electrode that is highly crystalline and on an amorphatized PLD silicon thin-film electrode, both the evolution of open-circuit potential as well as the estimated kinetic parameters do not vary, which indicates that they are intrinsic to lithiation and delithiation kinetics in silicon (*i.e.*, the rate of electrochemical reactions and not the rate of phase transformation is limiting). A sample comparison between the estimated kinetic parameters at 50% SOC in crystalline and amorphous silicon thin-film electrodes is given in Table 2. For sake of brevity, a detailed description of similar studies on amorphous Si thin-film electrodes is not described here.

*Closed-circuit simulation.* – The estimated thermodynamic and kinetic parameters were then used to predict experimental data under other conditions. The closed-circuit experiment at a constant applied current could be written as,

$$I_{app} = C_{dl}\frac{dV}{dt} + i_0\left\{\exp\left[\frac{\alpha_a F}{RT}(V-U)\right] - \exp\left[-\frac{\alpha_c F}{RT}(V-U)\right]\right\} \quad [10]$$

Ohmic and diffusion losses were ignored because the total current was small. Equation 14 was solved numerically using a finite-difference routine, and Figure 12 shows the simulation results for $I_{app}$ = 20.83 μA/cm$^2$. The model correctly predicts the potential offset during both lithiation and delithiation. The model does not predict the potential evolution on change in current (near z = 1), which requires a more accurate estimation of the capacitance. The model also deviates from features in the data seen at z = 0 and z = 0.6, both of which require a better open-circuit potential estimation. The departure at z = 1 could be related to the Si phase diagram. Figure 13 shows the simulation results corresponding to lithiation and delithiation at different rates. The model correctly predicts a lower offset potential for the lithiation reaction than that for the delithiation reaction as seen in Figure 5. Also, the offset potential decreases with decrease in the lithiation/delithiation rate (Figure 14). This is not apparent in the data shown in Ref. 5 and in Figure 5 because the practical rates of lithiation/delithiation are within an order of magnitude whereas the decrease in the potential offset becomes significant if the lithiation/delithiation rates were decreased by four to five orders of magnitude.

*Implications.* – Cell efficiency for the Li$_x$Si/cathode system for a given SOC can be written as,

$$\% \text{ Efficiency} = \left[1 - \frac{(V_c - V_a^l) - (V_c - V_a^d)}{(V_c - V_a^l)}\right] \times 100 \quad [11]$$

where $V_c$ is cathode potential, $V_a^l$ and $V_a^d$ are anode potentials *vs.* Li/Li$^+$ corresponding to lithiation and delithiation, respectively. The efficiency of the Li$_x$Si system approaches 100% as the offset between the lithiation and the delithiation reaction potentials (*i.e.*, $V_a^d - V_a^l$) approaches 0. Figure 14 shows the predicted efficiency as a function of the ratio of lithiation/delithiation current to the exchange current density. A cathode potential of 3.8 V *vs.* Li/Li$^+$ was used for this calculation. For a C/10 rate, the model predicts approximately 91% energy efficiency. The minimum overall cell-level energy-efficiency goal set by the Office of Vehicle Technologies for power-assist hybrid electric vehicles is 90% on a load profile with variable rates.[73,74] USABC's long-term cell-level efficiency goal for advanced-battery technologies is 80% for a C/3 discharge followed by a 6-hour charge.[75] The model predicts that high-energy efficiency could be obtained only by decreasing I/(aLi$_0$) by five orders of magnitude. Since $i_0$ is an inherent property of the Li$_x$Si system, the potential offset cannot be removed, and an increase in efficiency could be achieved only by increasing the surface area of the electrode also by a factor of 10$^5$. The model also shows that high rates are possible despite having a potential offset.

## Conclusions

The kinetics of lithiation and delithiation in silicon was studied on a model thin-film crystalline-silicon electrode. A Tafel equation is shown to account for the irreversibility between lithiation and delithiation exhibited by a well cycled electrode. The potential hysteresis between the lithiation and delithiation reactions at any given SOC exhibited by the Li$_x$Si system is shown to be caused by a very large kinetic resistance (small $i_0$). In conjunction with a model based on Tafel kinetics and double layer capacitance, data obtained from the open-circuit relaxation experiments is used to estimate the apparent transfer coefficients and the ratio of exchange current density to double-layer capacitance, all as a function of SOC for both lithiation and delithiation reactions. With these parameters, the model is shown to predict successfully the behavior of the system under constant-current lithiation and delithiation. The model also predicts that high-energy efficiency (or lowering the potential offset) could be obtained only by reducing $I/(aLi_0)$ by five orders of magnitude. In other words, a large increase in the surface area is required to enhance the energy efficiency of the silicon anodes. Overall, this simple model helps in understanding the limitations, and provides guidance to improving the performance of the Li$_x$Si system.

A methodology for the estimation of the side-reaction rate in this system is presented. The side-reaction rate kinetic parameters explain the open-circuit-potential relaxation at longer time scales. A more detailed understanding of the impact of the side reaction is needed to understand and minimize the cycle-to-cycle capacity loss observed in this system. Also, more understanding about the structural changes between lithiation and delithiation is needed. The thermodynamic and kinetic parameters estimated in this study are currently being incorporated into a transport model to understand the true rate-capability limitations in this system. In addition, stress-potential relationships in this system are being measured experimentally in our laboratory to understand how stress contributes to the potential offset.

# Tables

**Table 1.** Parameters used in the analysis of $Li_xSi$ electrode.

| Parameter | Value | Comments |
|---|---|---|
| A | 2.54 cm$^2$ | Measured |
| $\alpha_{side}$ | 0.5 | Assumed |
| $C_{dl}$ | 0.02 F/cm$^2$ | [a]Estimated |
| F | 96485 C/mol | Ref. 76 |
| $i_{0,side}$ | 7.5 x 10$^{-13}$ A/cm$^2$ | [a]Estimated |
| $i_0/C_{dl}$ | 4.23 nV/s | Estimated |
| R | 8.314 J/mol/K | Ref. 76 |
| T | 298 K | Measured |
| $U_{side}$ | 0.8 V vs. Li/Li$^+$ | Assumed |
| $V_c$ | 3.8 V | Assumed |

[a] – based on geometric area.

**Table 2.** Parameters estimated from open-circuit potential relaxation at 50% SOC from three different thin-film electrodes.

| | PLD Si | | Amorphous Si |
|---|---|---|---|
| Parameter | First cycle | Steady-state cycles | Steady-state cycles |
| $i_{0,a}/C_{dl}$ | 8.12 nV/s | 2.77 nV/s | 4.13 nV/s |
| $\alpha_a$ | 1.93 | 1.95 | 2.14 |

# List of Symbols

| | |
|---|---|
| $C_{dl}$ | double layer capacitance, F/cm$^2$ |
| $I_{app}$ | applied current density, A/cm$^2$ |
| $i_0$ | exchange current density, A/cm$^2$ |
| F | Faraday's constant, 96485 C/mol |
| L | film thickness, cm |
| T | temperature, K |
| t | time, s |
| U | equilibrium potential, V |
| V | potential, V |
| $V_0$ | potential at time zero, V |
| $V_c$ | cathode potential, V |
| $V_a^l$ | anode potential corresponding to lithiation, V |
| $V_a^d$ | anode potential corresponding to delithiation, V |
| z | State of charge |
| **Greek** | |
| $\alpha_a$ | apparent transfer coefficient corresponding to delithiation |
| $\alpha_c$ | apparent transfer coefficient corresponding to lithiation |
| $\alpha_{side}$ | apparent transfer coefficient corresponding to the side reaction |
| $\lambda$ | wavelength, nm |

φ   diameter, cm

## Acknowledgements

The authors gratefully acknowledge the financial support from the Assistant Secretary for Energy Efficiency and Renewable Energy, Office of Vehicle Technologies, the United States Department of Energy under contract no. DE-AC02-05CH11231.  The authors thank Dr. Robert Kostecki and Dr. Laurence J. Hardwick (LBNL) for help with Raman spectroscopy measurements;  Xiaojun Zhang and Dr. Samuel S. Mao (LBNL) for help with pulse-laser deposition; Dr. Keith D. Kepler (Farasis Energy, Inc.) for help with coin-cell fabrication; Xiangyun Song (LBNL) for help with electron microscopy and X-ray measurements.  The authors acknowledge the support of the National Center for Microscopy at LBNL, which is supported by the United States Department of Energy under contract no. DE-AC02-05CH11231.  The authors acknowledge the Microlab at the Department of Electrical Engineering in the University of California, Berkeley, for the fabrication of amorphous Si thin films.

# Figure Captions

Figure **1**: Scanning electron micrographs of a pulsed-laser-deposited silicon film. The inset shows the histogram of mean crystallite size estimated from the image on the left**. The scale bar on the left is 100 nm, and that on the right is 1 μm.**

Figure **2**: Raman spectrum obtained on a pulse-laser-deposited Si thin film is compared to that of a single-crystal (100) Si wafer. Spectra are shifted arbitrarily up the intensity axis for clarity. The peak at 521 cm$^{-1}$ is characteristic of a highly crystalline Si.

Figure **3**: Cell potential *vs.* capacity curves corresponding to lithiation and delithiation of pulse-laser-deposited Si thin-film electrode cycled at C/8 rate between 1.2 and 0.01 V *vs.* Li/Li$^+$. These curves were obtained on a well cycled cell. Inset: Cell potential *vs.* capacity curves corresponding to the first lithiation and delithiation of the same electrode.

Figure **4**: Lithiation and delithiation capacities for the first five cycles shown along with the respective cycling efficiency.

Figure **5**: Cell potential *vs.* capacity curves for different lithiation and delithiation rates (from C/30 to C/2) cycled between 0 and 50% SOC. **The electrode was cycled 10 times before conducting the experiment.**

Figure **6**: Relaxation of open-circuit potential (data) at 50% SOC during delithiation (upper curve) and lithiation (lower curve). The potential was changing even after 48 hours.

Figure **7**: Cell potential *vs.* capacity curves for lithiation and delithiation of a pulse-laser-deposited Si thin-film electrode cycled at a C/8 rate between 1.2 and 0.01 V *vs.* Li/Li$^+$ shown in (a) is corrected for side reaction, and the result is shown in (b).

Figure 8: Open-circuit potential at the end of a ten-hour relaxation experiment during lithiation (-■-) and delithiation (-●-) experiments as a function of SOC. The solid line corresponds to a lithiation/delithiation experiment between 1.2 and 0.01 V *vs.* Li/Li$^+$ at a constant current density corresponding to C/8, and the dotted line represents the potential used for model simulations (see text for details).

Figure **7**: Relaxation of cell potential recorded at open-circuit compared with model fits with and without side reaction. The points correspond to open-circuit-potential data (-■-) at 83% SOC during lithiation; the solid and the dotted lines respectively correspond to Marquardt-Levenberg fits with and without side reactions.

Figure **8**: Estimated main-reaction and side-reaction currents during open-circuit potential relaxation at 83% SOC during lithiation.

Figure **9**: Apparent transfer coefficients for delithiation ($\alpha_a$, -●-) and lithiation ($\alpha_c$, -■-) reactions at various SOCs estimated from the open-circuit-potential relaxation experiments.

Figure **10**: Values for $i_0/C_{dl}$ corresponding to lithiation (-■-) and delithiation (-●-) reactions at various SOCs estimated from the open-circuit-potential relaxation experiments.

Figure **11**: Potential *vs.* capacity for the first lithiation and delithiation between 1.2 and 0.01 V *vs.* Li/Li$^+$ is shown along with ten-hour open-circuit-relaxation data at various intervals. **The inset shows the open-circuit-potential evolution at 50% SOC along with the model fit based on equation 7.**

Figure **12**: Simulation (dashed line) and data (points) corresponding to a constant current ($I_{app}$ = 20.83 µA/cm$^2$) lithiation and delithiation between 0 and 1.2 V *vs.* Li.

Figure **13**: Simulated curves corresponding to different lithiation/delithiation rates ranging from C/2 to C/30. The offset potential decreases with decrease in the C rate.

Figure **14**: Combined potential offset (-■-) and percent efficiency (-●-) of lithiation/delithiation reactions for different C rates as predicted by the model.

# Figures

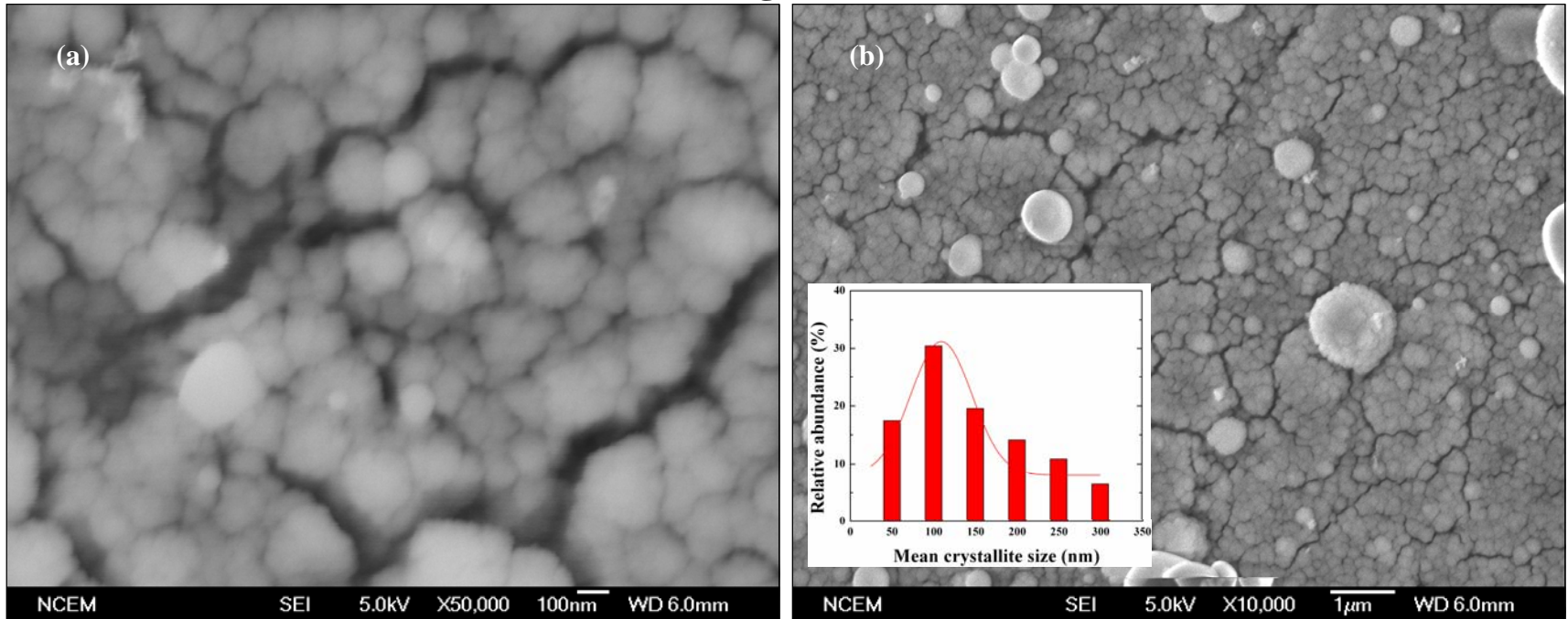

Figure 1: Scanning electron micrographs of a pulsed-laser-deposited silicon film. The inset shows the histogram of mean crystallite size estimated from the image on the left. The scale bar on the left is 100 nm, and that on the right is 1 μm.

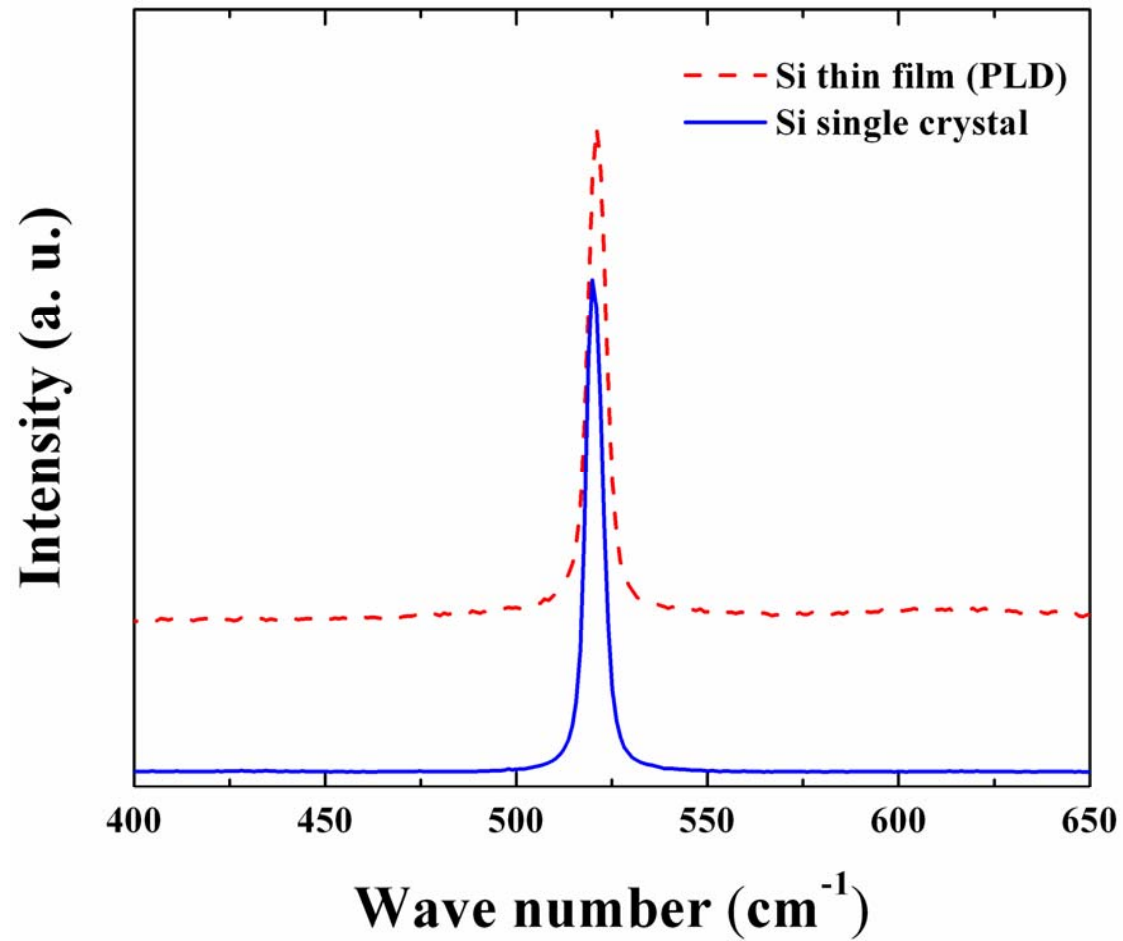

Figure 2: Raman spectrum obtained on a pulse-laser-deposited Si thin film is compared to that of a single-crystal (100) Si wafer. Spectra are shifted arbitrarily up the intensity axis for clarity. The peak at 521 cm$^{-1}$ is characteristic of a highly crystalline Si.

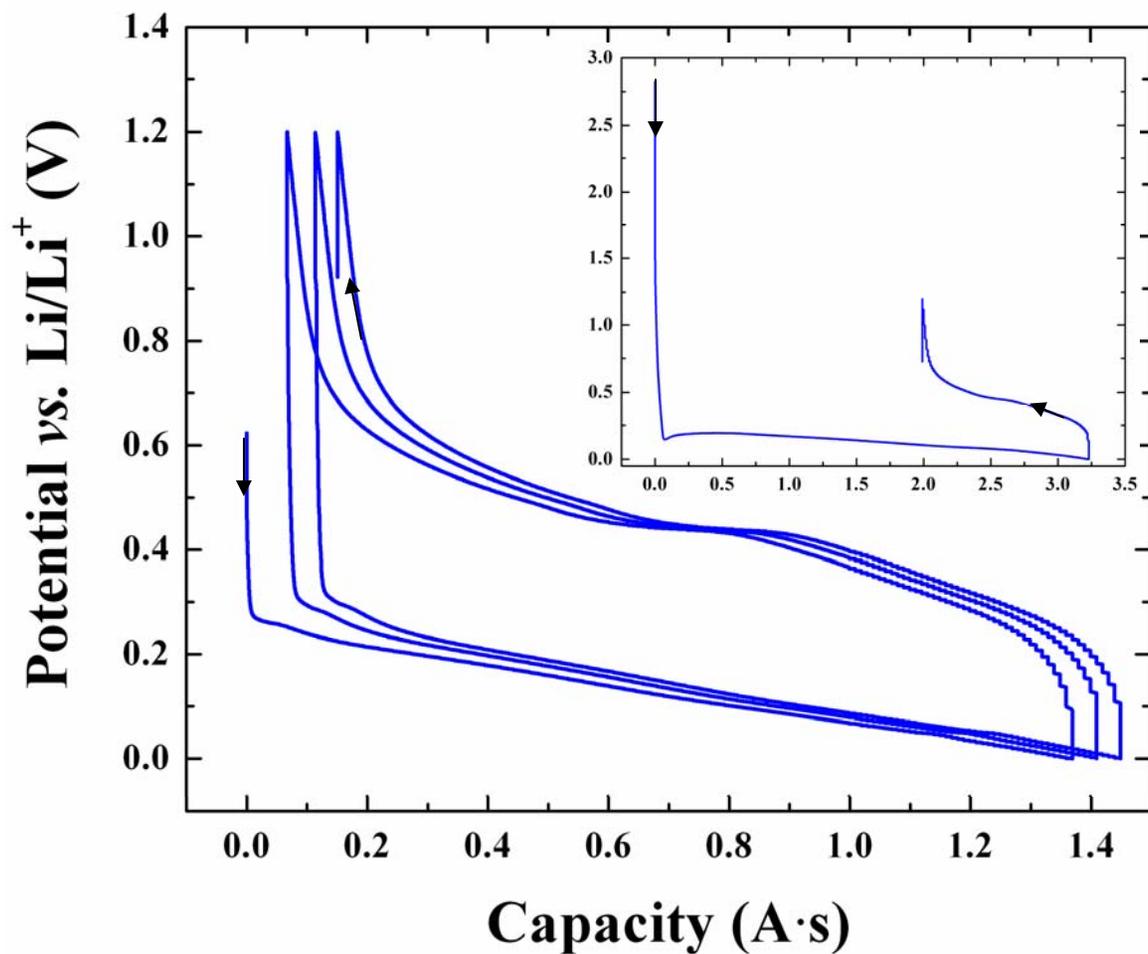

Figure 3: Cell potential *vs.* capacity curves corresponding to lithiation and delithiation of pulse-laser-deposited Si thin-film electrode cycled at C/8 rate between 1.2 and 0.01 V *vs.* Li/Li$^+$. These curves were obtained on a well cycled cell. Inset: Cell potential *vs.* capacity curves corresponding to the first lithiation and delithiation of the same electrode. The arrows in both figures indicate the cycling direction.

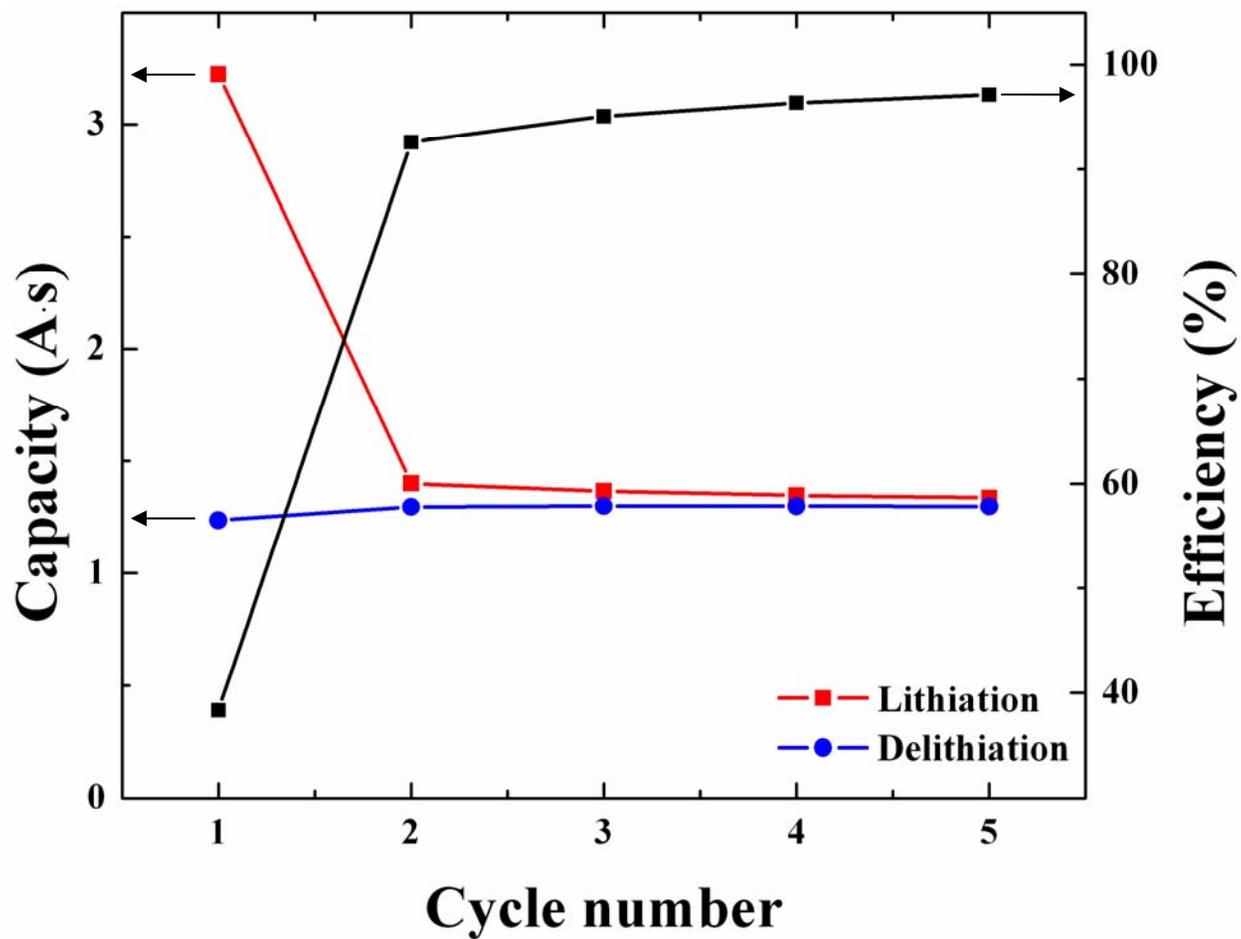

Figure 4: Lithiation and delithiation capacities for the first five cycles shown along with the respective cycling efficiency. The experiments were conducted at C/8.

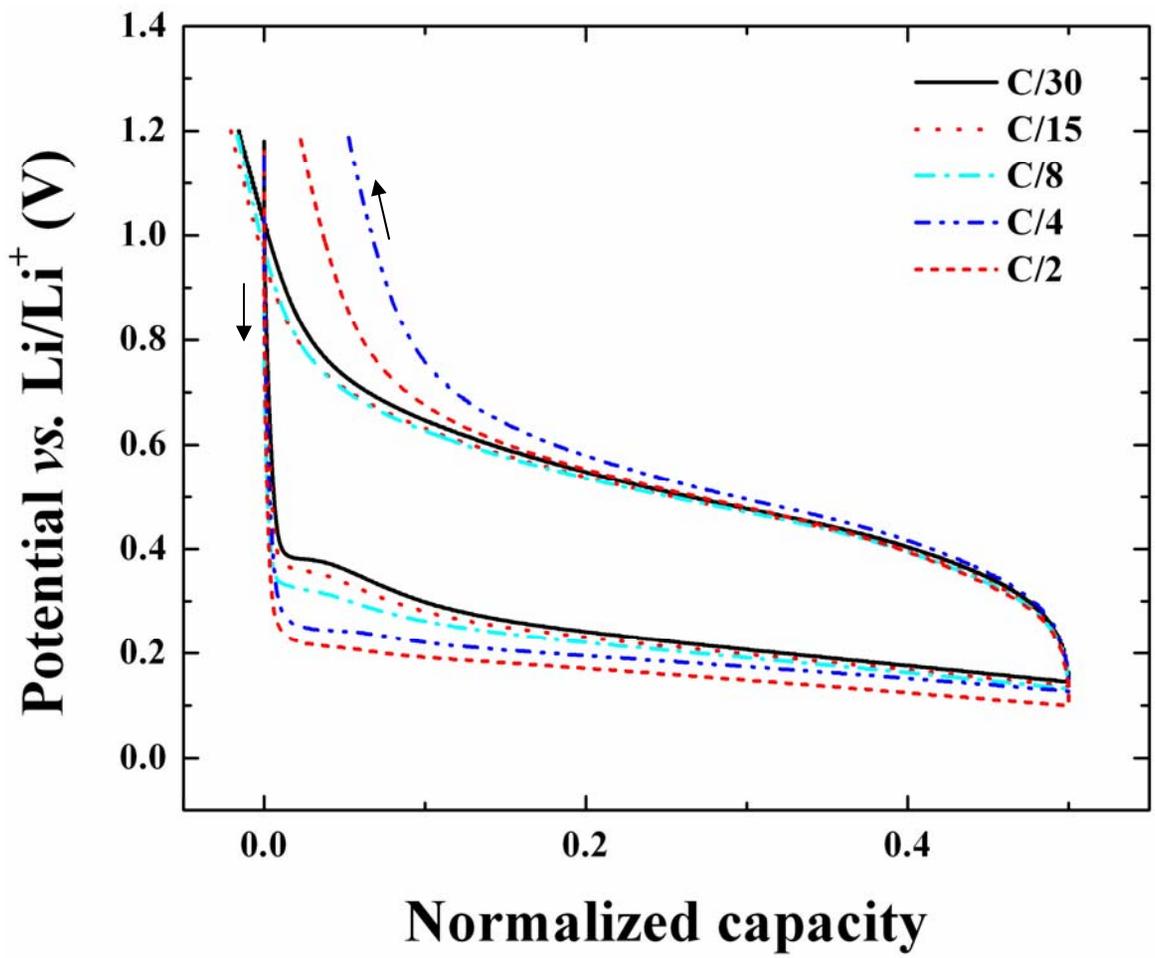

Figure 5: Cell potential *vs.* capacity curves for different lithiation and delithiation rates (from C/30 to C/2) cycled between 0 and 50% SOC. The electrode was cycled 10 times before conducting the experiment.

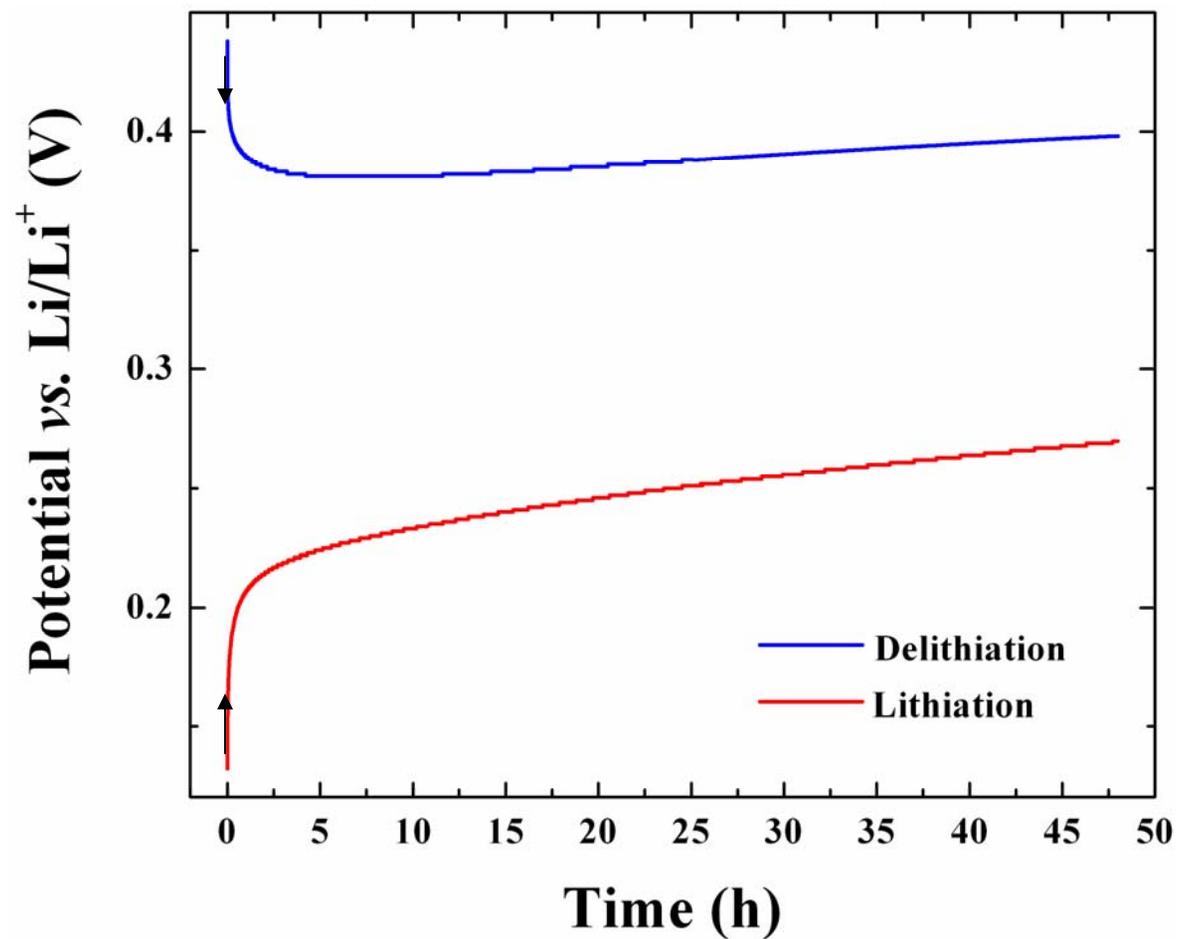

Figure 6: Relaxation of open-circuit potential (data) at 50% SOC during delithiation (upper curve) and lithiation (lower curve). The potential was changing even after 48 hours.

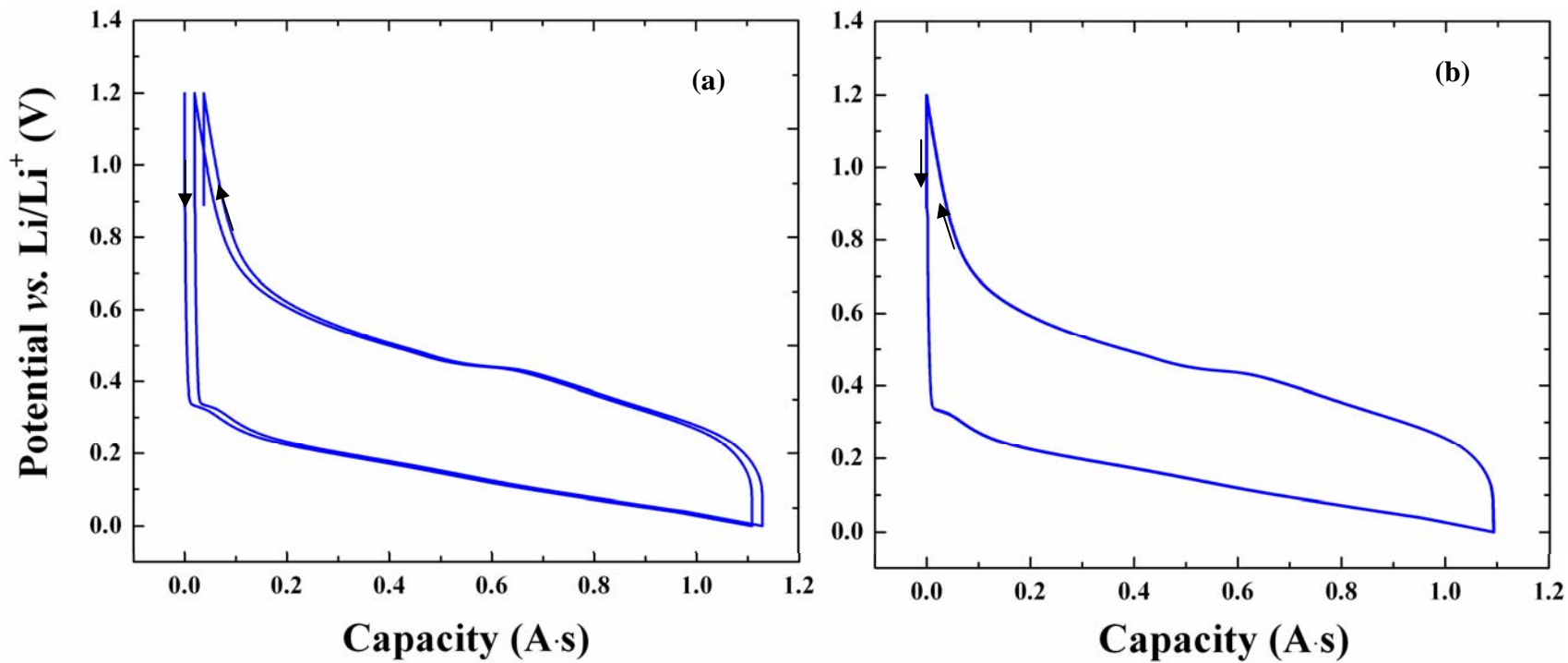

Figure 7: Cell potential *vs.* capacity curves for lithiation and delithiation of a pulse-laser-deposited Si thin-film electrode cycled at a C/8 rate between 1.2 and 0.01 V *vs.* Li/Li$^+$ shown in (a) is corrected for side reaction, and the result is shown in (b).

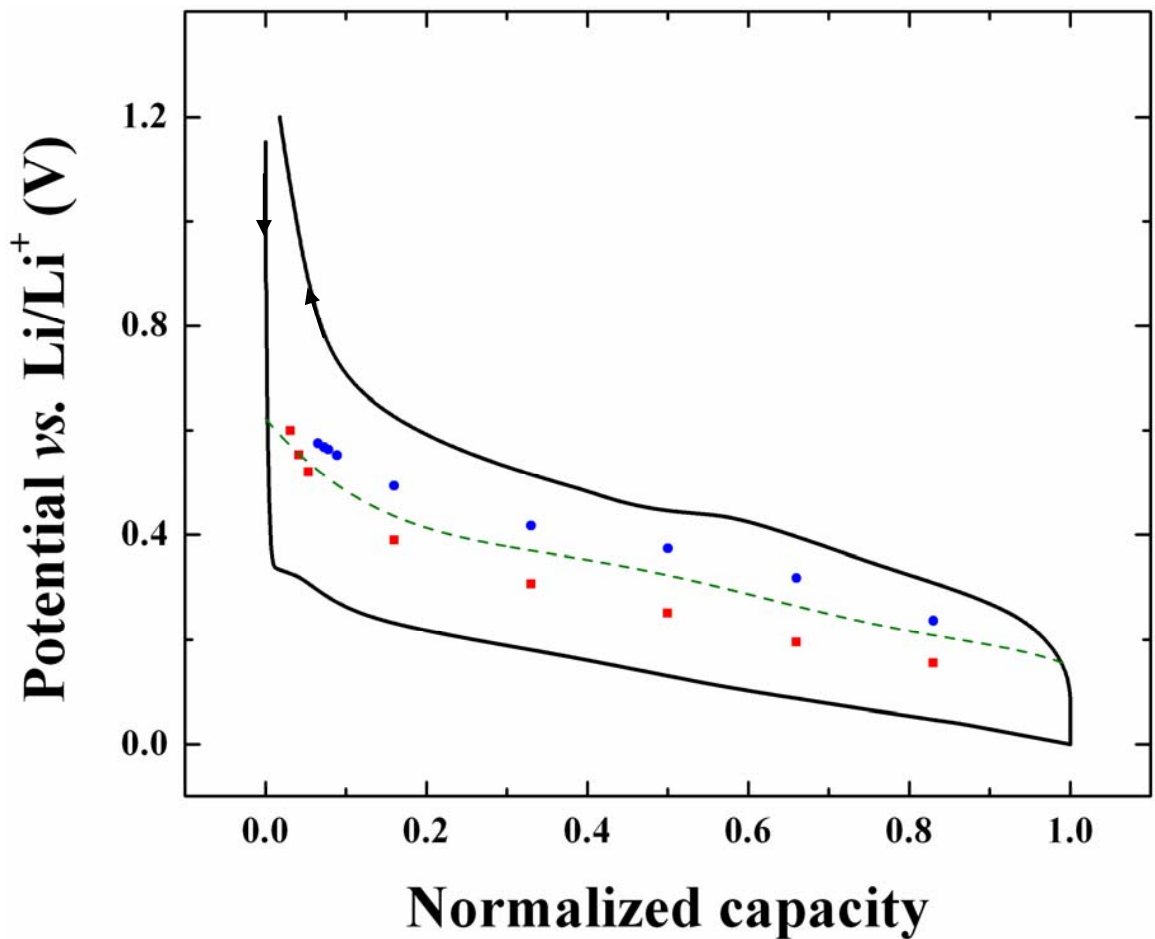

Figure 8: Open-circuit potential at the end of a ten-hour relaxation experiment during lithiation (-■-) and delithiation (-●-) experiments as a function of SOC. The solid line corresponds to a lithiation/delithiation experiment between 1.2 and 0.01 V *vs.* Li/Li$^+$ at a constant current density corresponding to C/8, and the dotted line represents the potential used for model simulations (see text for details).

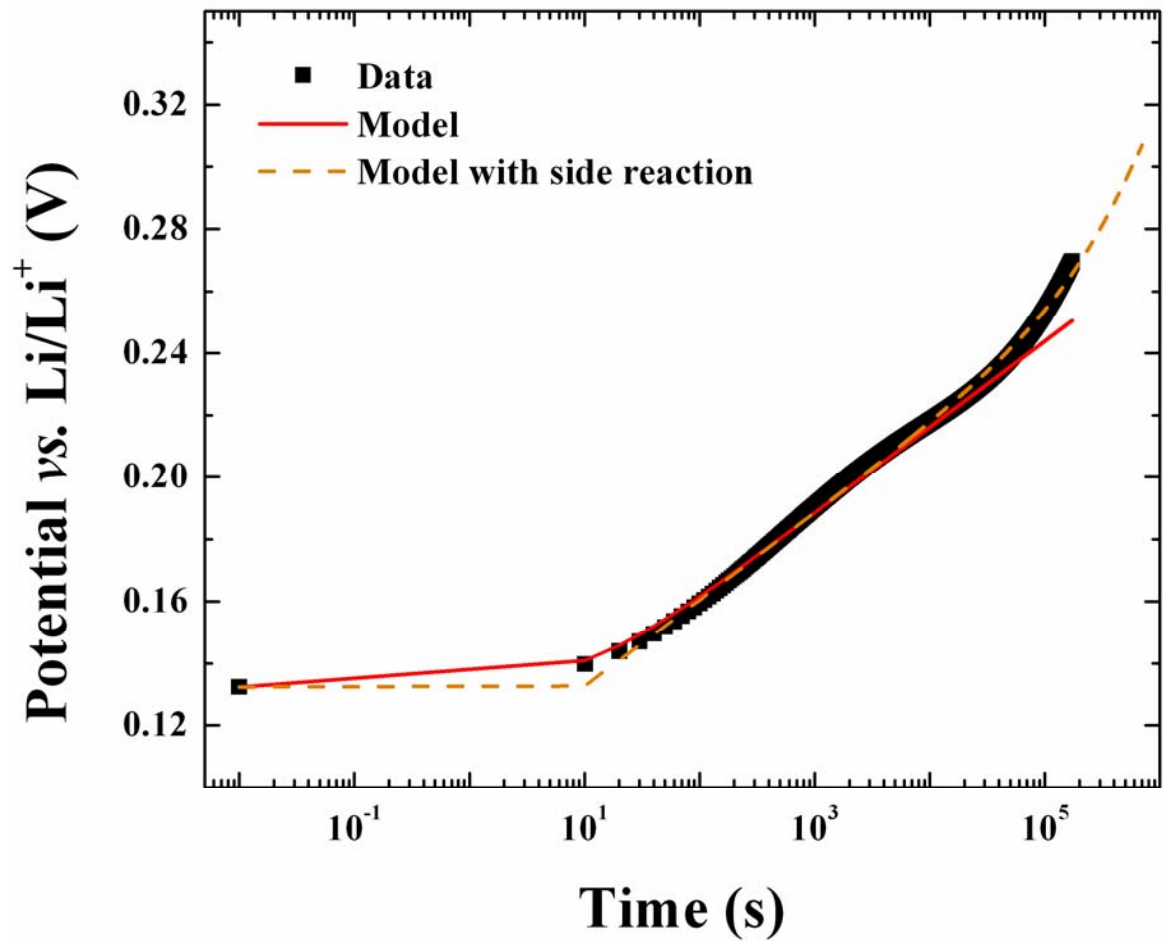

Figure 7: Relaxation of cell potential recorded at open-circuit compared with model fits with and without side reaction. The points correspond to open-circuit-potential data (-■-) at 83% SOC during lithiation; the solid and the dotted lines respectively correspond to Marquardt-Levenberg fits with and without side reactions.

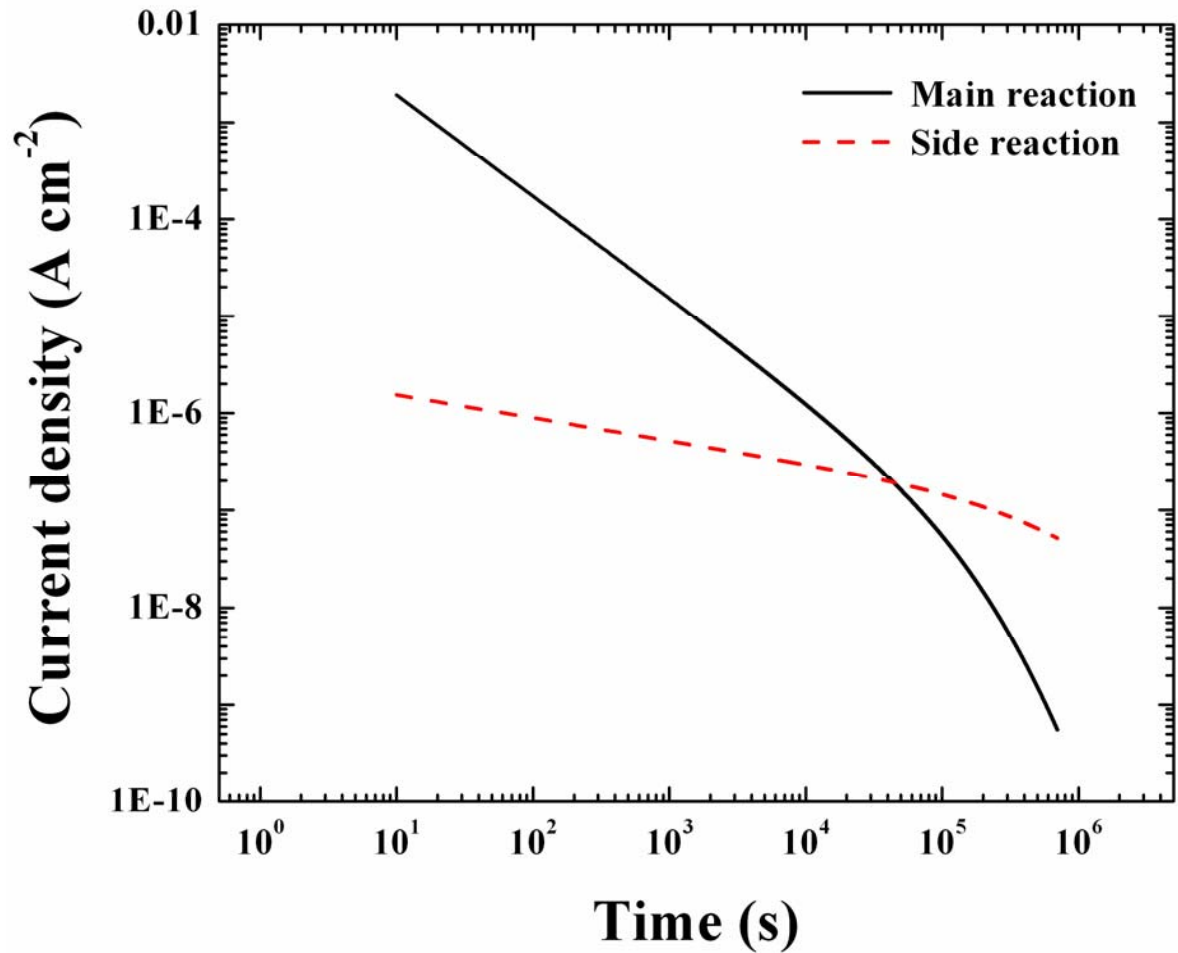

Figure 8: Estimated main-reaction and side-reaction currents during open-circuit potential relaxation at 83% SOC during lithiation.

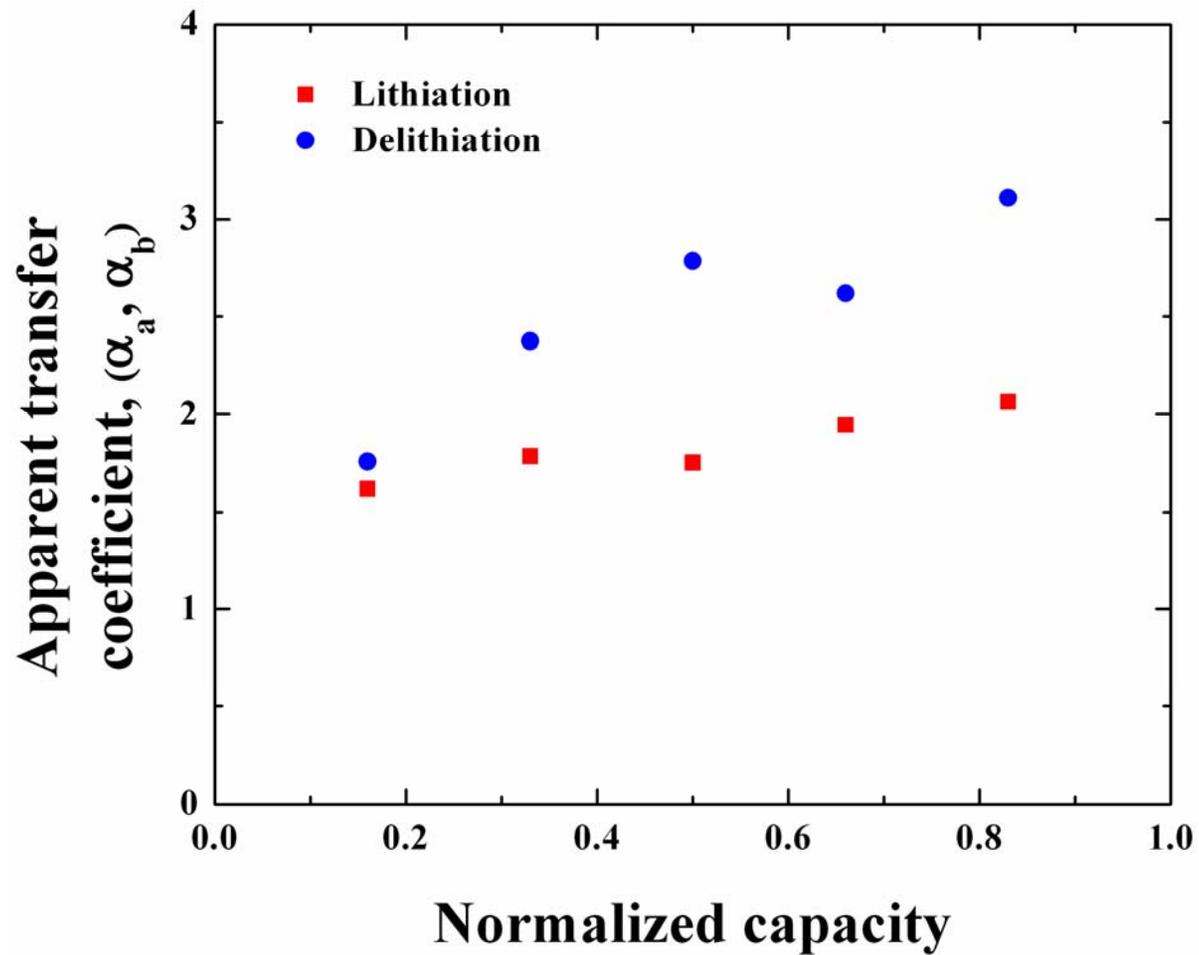

Figure 9: Apparent transfer coefficients for delithiation ($\alpha_a$, -●-) and lithiation ($\alpha_c$, -■-) reactions at various SOCs estimated from the open-circuit-potential relaxation experiments.

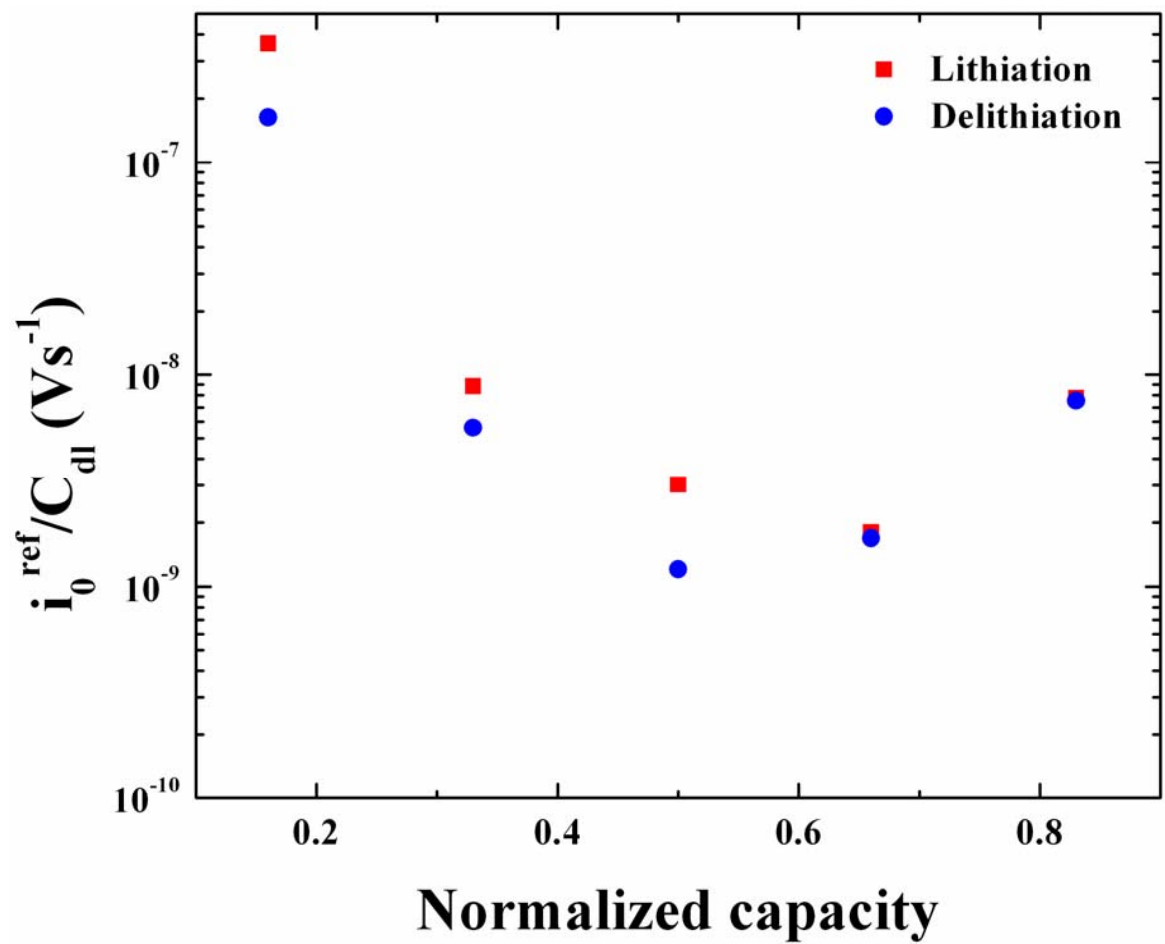

Figure 10: Values for $i_0/C_{dl}$ corresponding to lithiation (-■-) and delithiation (-●-) reactions at various SOCs estimated from the open-circuit-potential relaxation experiments.

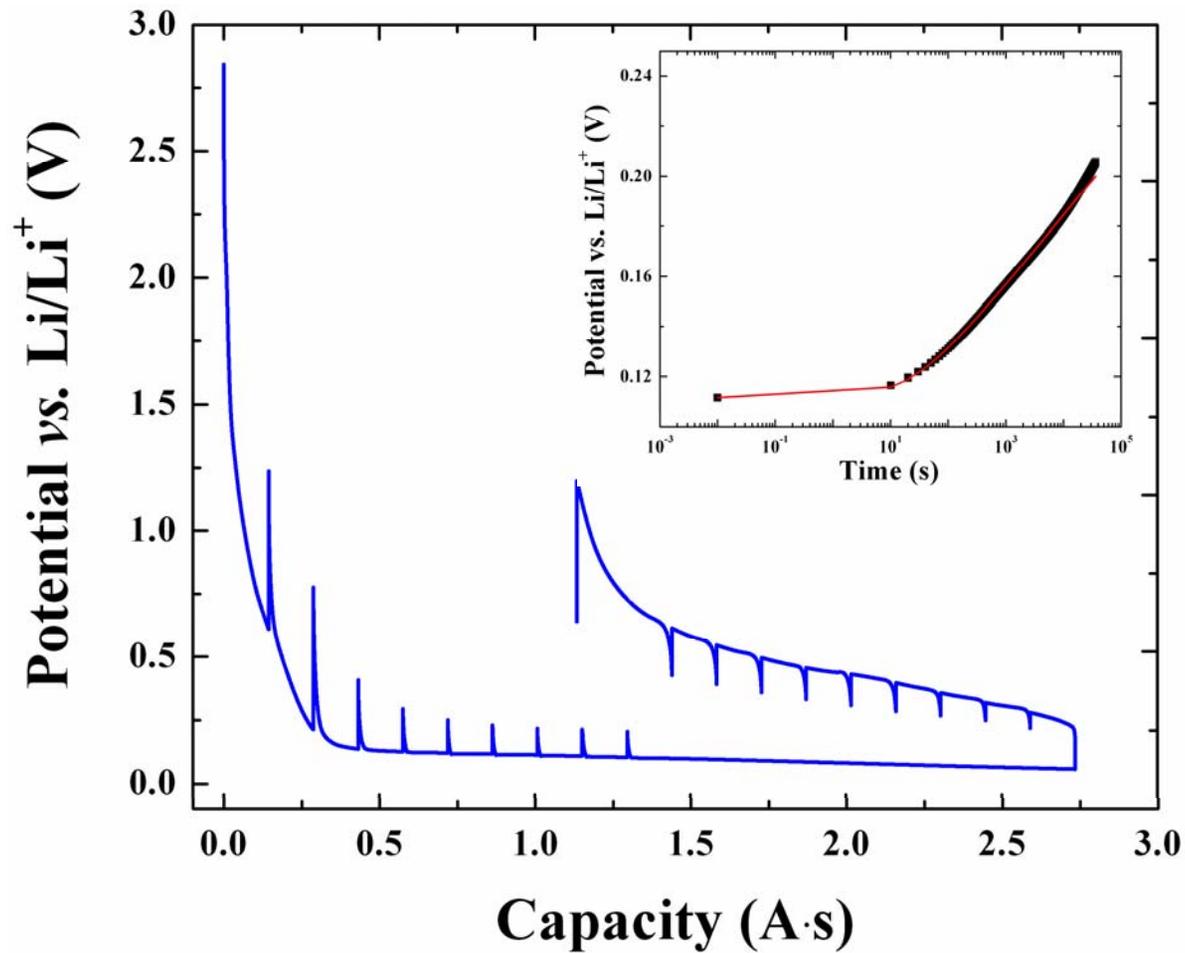

Figure 11: Potential *vs.* capacity for the first lithiation and delithiation between 1.2 and 0.01 V *vs.* Li/Li$^+$ is shown along with ten-hour open-circuit-relaxation data at various intervals. The inset shows the open-circuit-potential evolution at 50% SOC along with the model fit based on equation 7.

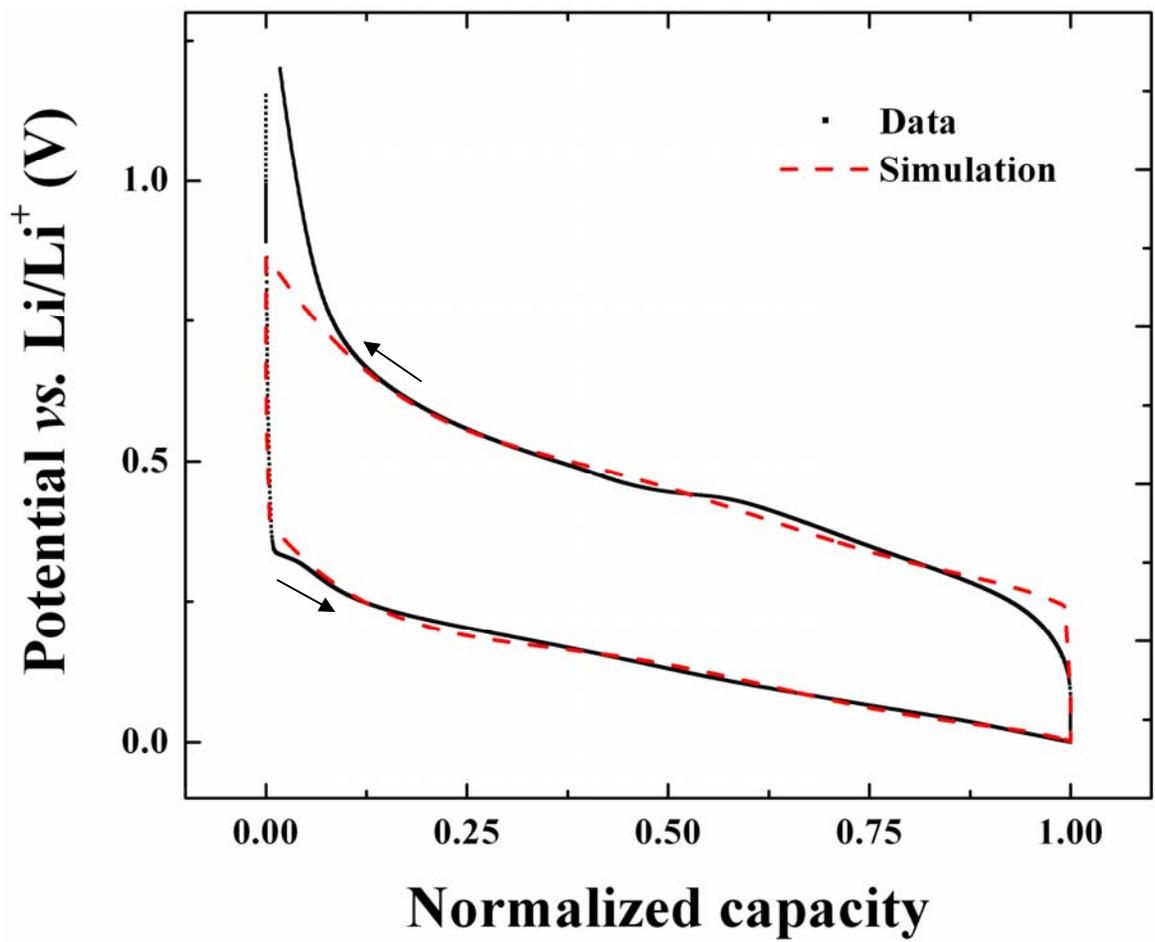

Figure 12: Simulation (dashed line) and data (points) corresponding to a constant current ($I_{app}$ = 20.83 µA/cm$^2$) lithiation and delithiation between 0 and 1.2 V *vs.* Li.

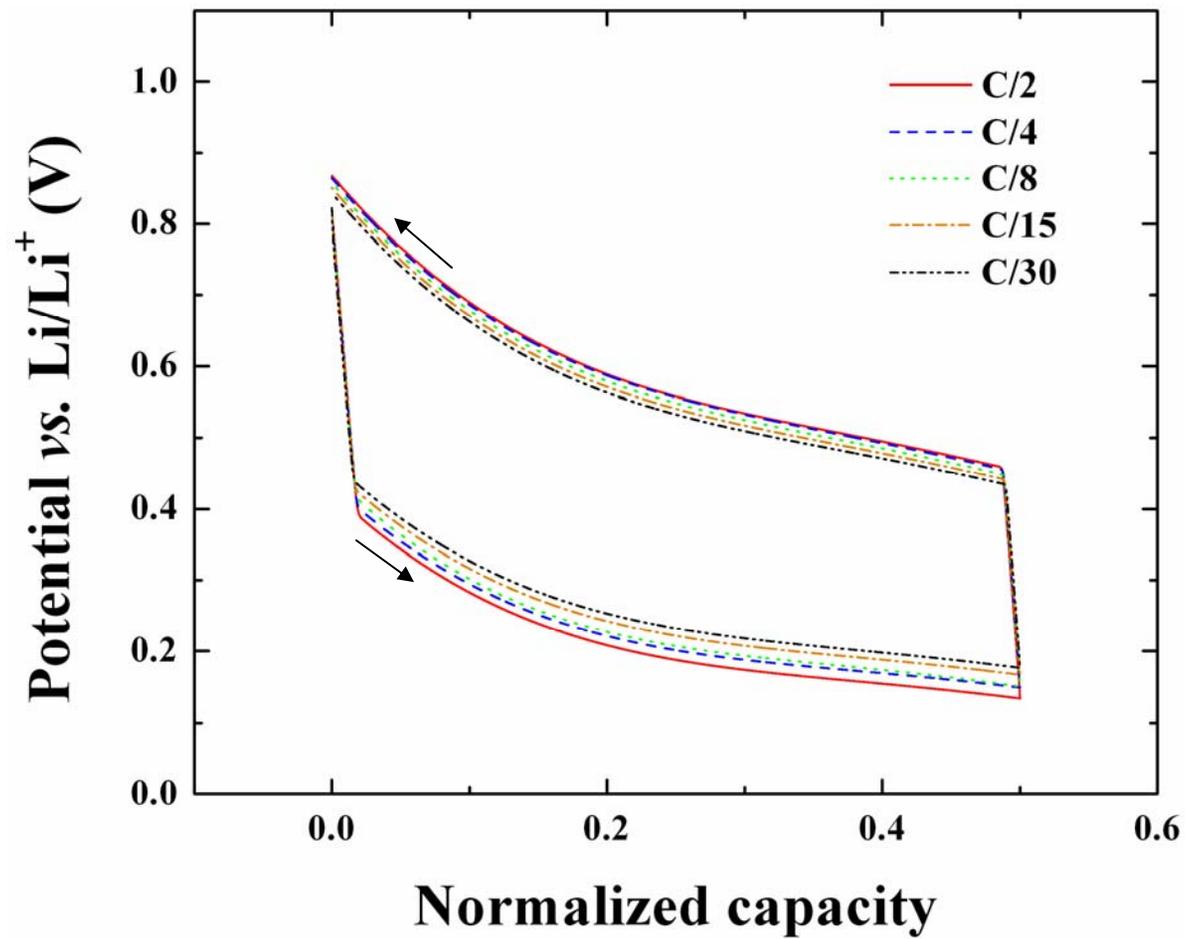

Figure 13: Simulated curves corresponding to different lithiation/delithiation rates ranging from C/2 to C/30. The offset potential decreases with decrease in the C rate.

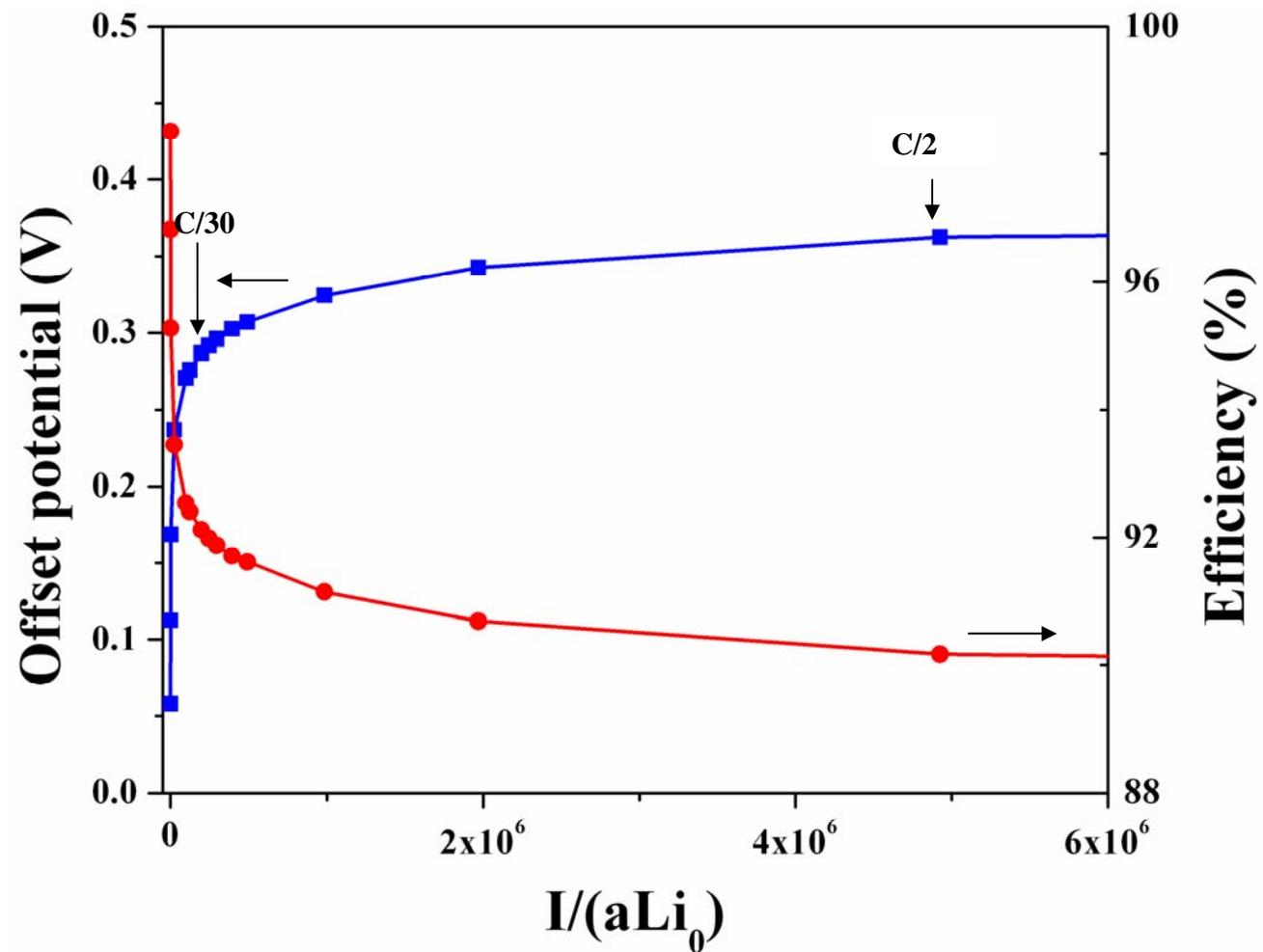

Figure 14: Combined potential offset (-■-) and percent efficiency (-●-) of lithiation/delithiation reactions for different C rates as predicted by the model.